\documentclass[seceq]{ptptex}

\usepackage{graphicx}
\usepackage{tabularx}

\title{
The statistical relationship between product life cycle and repeat 
purchase behavior in convenience stores}

\author{
Takayuki \textsc{Mizuno}$^{1,}$\footnote{E-mail: mizuno@ier.hit-u.ac.jp} 
and Misako \textsc{Takayasu}$^{2}$
}

\inst{
$^1$Institute of Economic Research, Hitotsubashi University, 2-1 Naka, Kunitachi 
City, Tokyo 186-8603, Japan
\\
$^2$Department of Computational Intelligence {\&} Systems Science, 
Interdisciplinary Graduate School of Science {\&} Engineering, Tokyo 
Institute of Technology, 4259 Nagatsuta-cho, Midori-ku, Yokohama 226-8502, 
Japan
}

\abst{
The density function of product life cycles in convenience stores is found 
to follow the Weibull distribution. To clarify the parameters that determine 
these life cycles, we introduce the conditional market share---defined as 
the probability that a product is selected by customers only if it had been 
previously purchased---and the market share without any conditions. The 
product life cycle is more strongly correlated with the conditional market 
share of the product than with the latter type of market share. 
}

\begin{document}

\maketitle

\section{Introduction}

Over the last decade, physicists and economists engaged in the field of 
econophysics have actively studied financial markets by analyzing large 
amounts of financial data using techniques of statistical physics \cite{rf:1,rf:2}. 
The results of these studies have often been applied to financial 
businesses. In order to investigate economic phenomena by using such an 
empirical approach, there is a need for a large number of datasets 
demonstrating these phenomena. Therefore, experts in the field first 
attempted to use commonly available financial datasets. Recently, many 
datasets except those on the financial markets have become available owing 
to the developments in information technology. We have studied the 
statistical laws governing spending by an individual in convenience stores 
by analyzing about 100 million receipts. The expenditure patterns exhibit 
violent fluctuations observed in nonequilibrium situations \cite{rf:3,rf:4}. In the 
past, D. Sornette and F. Deschatres et al. presented an extensive study on 
the typical patterns of the time series of book sales by using a large 
dataset of book sales obtained from Amazon.com \cite{rf:5,rf:6}. Through our study, we 
explain more general economic phenomena.

In this paper, we analyze scanner data comprising recorded customer history 
in the convenience store chain of ``am/pm Japan Co. Ltd.'' Such scanner data 
are mainly investigated in field of marketing. However, statistical laws 
governing customer purchase behavior have seldom been investigated 
scientifically because typically, the focus when using such data is on 
identifying the marketing strategies companies that can apply to efficiently 
obtain profits. In this paper, we clarify the statistical laws governing 
product life cycle and consumer repeat purchase behavior.

B. Klumb et al. outlined a framework for increasing the likelihood of 
efficient new product introductions. Using Nielsen/BASES data on 850 new 
products introduced in the US, they confirmed that repeat customers are 
important for increasing the sales of a new product \cite{rf:7}. In addition, the 
dependence on the diffusion of new products on trial-and-repeat purchases was 
investigated through model analyses \cite{rf:8,rf:9,rf:10}. Through our study, we show that 
there are correlations between the life cycle of each product and the 
percentage of repeat buyers of a product.

This paper is organized as follows. We first introduce the scanner data. 
Next, we focus on the statistical laws governing product life cycle in 
convenience stores. Most products are replaced with new products in only a 
few months. For example, 484 kinds of rice ball were released in stores from 
2004 to 2007. However, about 70{\%} of them disappeared from stores within 2 
months. We investigate the product life cycle on the basis of a survival 
analysis, which deals with the death of biological organisms and failure of 
mechanical systems \cite{rf:11}; survival analysis is also known as reliability 
analysis in engineering and duration analysis in economics or sociology. 
Next, we observe the repeat purchase behavior of customers. We clarify the 
probability structure determining the continued purchase of a product by a 
customer. Then, we show that a product life cycle depends on the probability 
structure. Finally, we discuss why the life cycle depends on repeat purchase 
behavior.

\section{Dataset}

In this study, we analyzed the POS database that recorded the customer 
purchase history for the convenience store chain under consideration, namely, 
am/pm Japan Co. Ltd. This company operates around 1,000 stores in Japan. The 
stores typically sell drinks, magazines, prepackaged foods such as rice 
balls and lunchboxes, toiletries, cigarettes, daily necessities, and 
miscellaneous goods. 

In recent years, the use of electronic money has increased in Japan. One of 
the more famous modes of making electronic payments in the country is 
through a service called ``Edy,'' provided by bitWallet, Inc. Edy is a 
prepaid smart card that is rechargeable and contactless and can be used to 
make payments in places such as convenience stores. Every Edy card has a 
unique number called the Edy ID. When a payment is made using an Edy card, 
the Edy ID is recorded in a POS database that is stored on a server. By 
accessing the Edy ID on the database, details of a purchase such as the 
receipt number, date of purchase, time of purchase, and Japan article number 
(JAN), which is specific to each product, can be retrieved.

To investigate product life cycles and consumer repeat purchase behaviors, we 
used 46,312,663 receipts of customers who made purchases using their Edy 
cards in am/pm Japan Co. Ltd. stores from December 26, 2003, to December 31, 
2007.

\begin{figure}[htbp]
\centerline{\includegraphics[width=3.97in,height=3.06in]{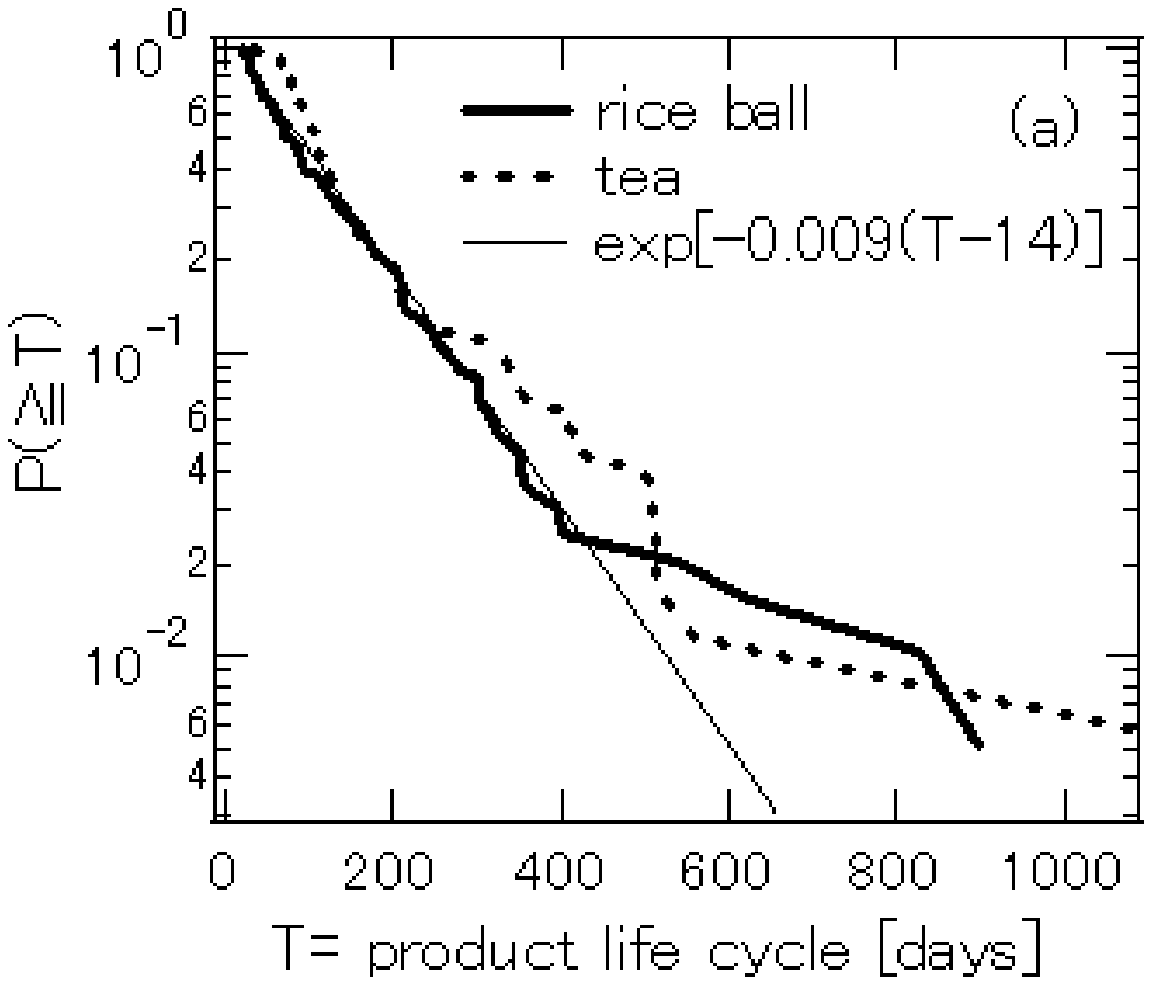}}
\label{fig1a}
\centerline{\includegraphics[width=3.90in,height=3.08in]{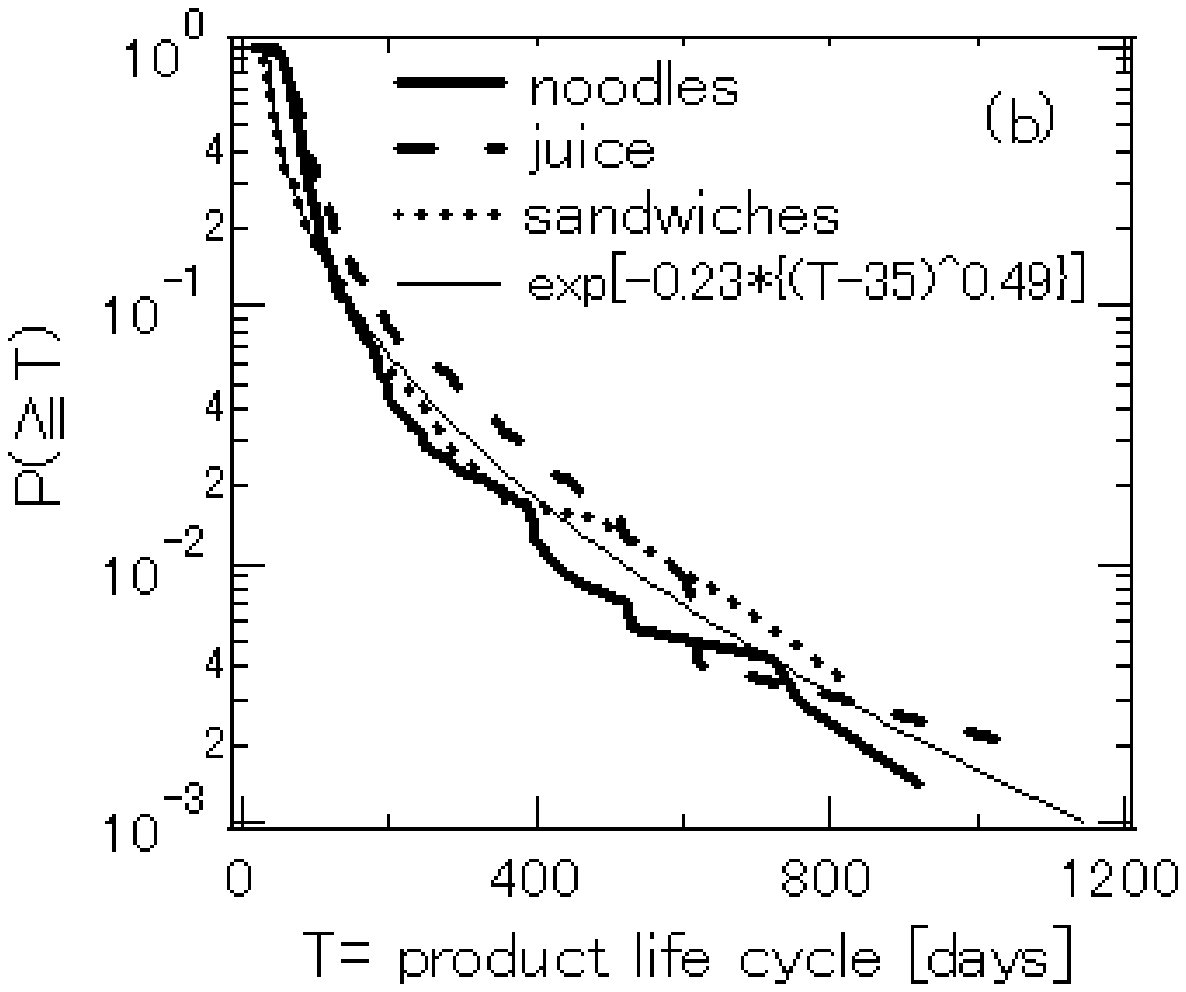}}
\label{fig1b}
\caption{Cumulative probability density functions of the product life 
cycle. (a) Rice ball and plastic bottle of green tea. (b) Instant noodles, 
plastic bottle of juice, and sandwiches. The lines show an exponential 
distribution, $P\left( {\ge T} \right)=e^{-0.009\left( {T-14} \right)}$, and 
a Weibull distribution, $P\left( {\ge T} \right)=e^{-0.23\left( {T-35} 
\right)^{0.49}}$.}
\end{figure}

\begin{table}[h]
\caption{AIC for the probability density function of the product 
life cycle. We approximate the function of each category by an exponential 
distribution and a Weibull distribution with coefficient $c$ of Eq.(3-1)}
 \begin{center}
  \begin{tabular}{l|lll}
    \hline
                      			& AIC of exponential $(c = 1)$ 	& AIC of Weibull	& $c$	\\
    \hline
    Rice ball       			& 967   			& 968       		& 0.97	\\
    Plastic bottle of green tea		& 745	  			& 742      		& 0.92	\\
    Sandwiches    			& 1005  			& 848     		& 0.49	\\
    Plastic bottle of juice		& 1846  		 	& 1725    		& 0.65	\\
    Instant noodles   			& 2504  			& 2372   		& 0.65	\\
    \hline
  \end{tabular}
 \end{center}
\end{table}

\section{Probability density function of product life cycles}

The life cycles of products in convenience stores are surprisingly short. In 
Fig.1, we present a cumulative probability function of the product life 
cycles. The vertical axis, $P\left( {\ge T} \right)$, shows the probability 
of finding a product with a life cycle longer than $T$ days. The product life 
cycles are approximated by the Weibull distribution, as follows:

\begin{equation}
P(\ge T)=e^{-a\left( {T-b} \right)^c}
\end{equation}

\noindent
where $a$ is the scale parameter and $c$ is the shape parameter of the 
distribution, and $b$ represents the minimum life cycle of each product 
category. The Weibull distribution is often used in the field of life data 
analysis due to its flexibility in cases such as the investigation of the 
failure rate of machine parts. If the failure rate decreases over time, $c < 1$. If 
the failure rate is constant over time, $c = 1$. If the failure rate increases over 
time, $c > 1$. When $c = 1$, the Weibull distribution becomes an exponential 
function. For each category, we measure the goodness of fit of an 
approximated distribution by using Akaike's information criterion (AIC). We 
approximate the product life cycles by the exponential distribution,$P\left( 
{\ge T} \right)=e^{-a\left( {T-b} \right)^c}_{ }$with $c = 1$, and the Weibull 
distribution,$P\left( {\ge T} \right)=e^{-a\left( {T-b} \right)^c}$. Table I 
shows the AICs of the two distributions. 

First, with regard to the products considered, we focus on rice balls and 
plastic bottles of green tea. The product life cycles for each category are 
approximated using the exponential function as shown in Fig.1(a). The AIC of 
Weibull distribution shows almost the same value as the AIC of exponential 
distribution as shown in Table I. Therefore, the life cycles of the 
categories follow Poisson processes because $c\cong 1$ in Eq.(3-1). 

Next, we also investigate instant noodles, plastic bottles of juice, and 
sandwiches. As shown in Fig.1(b), the product life cycles are approximated by 
the Weibull distribution with $c < 1$. The AIC of the Weibull distribution is 
sufficiently smaller than the AIC of exponential distribution as shown in 
Table I. Hence, $c < 1$ is a significant parameter. We cannot approximate the life 
cycles of these categories through pure Poisson processes because the 
probability that a product is discontinued depends on the number of days 
over which the product has been sold.

\section{Weak correlation between the life cycle and the market share 
without any conditions}

\begin{table}[h]
 \caption{The correlation coefficient between the product life cycle 
$(log T )$ and $P_k (N_{i,t} =1)$, $(log T )$, and $P_k (N_{i,t} =1\vert N_{i,t-1} 
=1)$. The market share with a condition, $P_k (N_{i,t} =1\vert N_{i,t-1} 
=1)$, is defined by the conditional probability that product $k$ is 
consistently chosen. Market share, $P_k (N_{i,t} =1)$, is defined by 
probability that product $k$ is chosen without any conditions.}
 \begin{center}
  \begin{tabular}{l|ll}
    \hline
                      			& $P_{k}(1)$ & $P_{k}(1\vert1)$	\\
    \hline
    Rice ball       			& 0.27   & 0.61       	\\
    Plastic bottle of juice  		& 0.37   & 0.60      	\\
    Sandwiches    			& 0.38   & 0.54     	\\
    Plastic bottle of green tea		& 0.38   & 0.44    	\\
    Instant noodles   			& 0.31   & 0.34   	\\
    \hline
  \end{tabular}
 \end{center}
\end{table}

We considered the parameters that determined product life cycles. We first 
focused on the product market share that is estimated by the probability, 
$P_k (N_{i,t} =1)$, that the product $k$ is bought without any conditions in a 
category that includes the product $k$. Here, $N_{i,t} = 1$ when a customer $i$ buys 
product $k$ in his $t$-th purchase in the category. Fig.2 displays the time series 
of the market shares of two rice ball products. The market share of product 
$A$ was always higher than that of product $B$. However, product $A$ was discontinued 
after about 90 days and product $B$ was continued for over 150 days. 

We indicate the relationship between the product life cycle and the market 
share, $P_k (N_{i,t} =1)$, in the case of rice balls, as depicted in Fig.3. 
We estimated the average market share for each life cycle. When a life cycle 
is longer than about 200 days, it depends on the market share. However, we 
found that the life cycle is almost independent of the market share in the 
case of products with short life cycles. About 70{\%} of the products 
disappear from stores within 2 months of being released for sale. Therefore, 
for most products, it is difficult to extrapolate the life cycle from the 
market share. With regard to the several popular kinds of products available 
in convenience stores in Japan, we present the correlations between the 
product life cycle and the market share, $P_k (N_{i,t} =1)$, by using a 
cross-correlation function in Table II. Generally, the correlations are 
weakly independent of the type of product.

\section{Analysis of repeat purchase behavior using conditional 
probability}

In order to investigate the difference between products $A$ and $B$ in Fig.2, we 
introduced a conditional market share, $P_k (N_{i,t} =1\vert N_{i,t-1} =1)$, 
which expresses the probability that a product $k$ is selected by a customer 
$i$ only if it had been his ($t-1$ )-th purchase. With many products, we find that 
the conditional market share is much higher than the market share without 
any conditions, $P_k (N_{i,t} =1)$, as shown in Table III. Thus, customers 
exhibit characteristic behaviors during repeat purchases.

In Table III, we observe the market share with some conditions; this is defined 
by the probability $P_k (N_{i,t} =1\vert N_{i,t-1} =1,N_{i,t-2} =1,\cdots 
,N_{i,t-\tau } =1)$ that product $k$ is chosen by customer $i$ who has 
consistently purchased it $\tau $ times. We do this to clarify the 
characteristics of repeat purchase behavior. The probability that product 
$k$ is chosen increases depending on its $\tau \mbox{-th}$ successive purchase. 
This characteristic is found in both products $A$ and $B$. In the case of product 
$B$, the probability that product $B$ is chosen without conditions, $P_B (N_{i,t} 
=1)$, is 0.026. However, the conditional probability, $P_B (N_{i,t} =1\vert 
N_{i,t-1} =1)$, that a customer who bought product $B$ the last time $(t-1)$ chooses 
product $B$ at time $t$ is 0.408. Furthermore, the conditional probability of a 
customer who bought product $B$ over $\tau =4$ successive times, $P_B (N_{i,t} 
=1\vert N_{i,t-1} =1,N_{i,t-2} =1,\cdots ,N_{i,t-\tau } =1)$, is higher than 
0.8. This characteristic of the conditional market share indicates that a 
customer gradually develops a sticky preference for the product by 
consistent purchase.

\begin{figure}[htbp]
\centerline{\includegraphics[width=4.04in,height=2.38in]{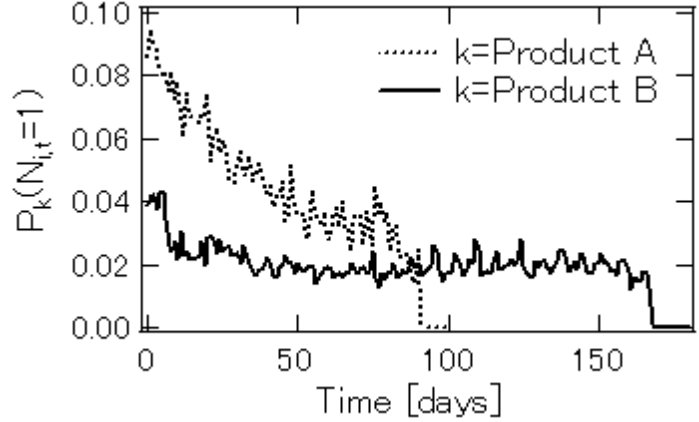}}
\label{fig2}
\caption{Time series of the market share, $P_k (N_{i,t} =1)$, of rice 
ball products $A$ and $B$. Product $A$ and $B$ were released for sale on February 22, 
2005, and December 12, 2006, respectively. Time = 0 indicates the release 
date.}
\end{figure}

\begin{figure}[htbp]
\centerline{\includegraphics[width=3.88in,height=2.83in]{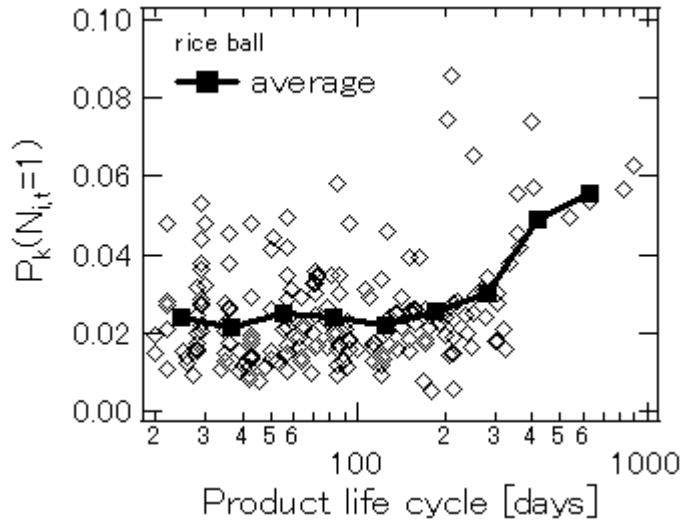}}
\label{fig3}
\caption{Relationship between product life cycle and market share, 
$P_k (N_{i,t} =1)$. The curve represents the average market share for each 
life cycle.}
\end{figure}

\begin{table}[h]
 \caption{Conditional market shares of product $A$ and $B$. Market share 
with conditions, $P_k (N_{i,t} =1\vert N_{i,t-1} =1,N_{i,t-2} =1,\cdots 
,N_{i,t-\tau } =1)$, expresses the conditional probability that product $k$ is 
chosen by customer $i$ who has consistently purchased it $\tau $ times. $N_{i,t} 
=0$ when customer $i$ does not buy product $k$ in his $t$-th purchase. Here, $P_k 
(N_{i,t} =1)$, is the probability that product $k$ is chosen in $t$-th purchase of 
rice balls without any conditions. The error term is defined by 
$\frac{1}{2\sqrt n }$, where $n$ is the number of data points.}
 \begin{center}
  \begin{tabular}{l|ll}
    \hline
    			&$k = $Product $A$      &$k = $Product $B$ \\
    \hline
    $P_{k}(1)$		&0.067 $\pm$ 0.001&	0.026 $\pm$ 0.001\\
    $P_{k}(1\mid1)$ 	&0.310 $\pm$ 0.004&	0.408 $\pm$ 0.004\\
    $P_{k}(1\mid11)$ 	&0.556 $\pm$ 0.009&	0.657 $\pm$ 0.006\\
    $P_{k}(1\mid111)$ 	&0.685 $\pm$ 0.012&	0.763 $\pm$ 0.007\\
    $P_{k}(1\mid1111)$ 	&0.763 $\pm$ 0.013&	0.823 $\pm$ 0.007\\
    $P_{k}(1\mid11111)$ 	&0.808 $\pm$ 0.014&	0.853 $\pm$ 0.007\\
    $P_{k}(1\mid111111)$ 	&0.852 $\pm$ 0.014&	0.865 $\pm$ 0.009\\
    \hline
  \end{tabular}
 \end{center}
\end{table}

\section{Relationship between product life cycle and conditional market 
share}

As shown in Table III, product $A$ with a short life cycle shows a low 
conditional market share, $P_A (N_{i,t} =1\vert N_{i,t-1} =1)=0.310$. 
However, product B with a long life cycle shows a high conditional market 
share, $P_B (N_{i,t} =1\vert N_{i,t-1} =1)=0.408$. On the basis of this 
result, we can point out the possibility that repeat purchase behavior 
determines the product life cycle. In order to quantify the degree of repeat 
purchase behavior for each product, we only focused on the conditional 
market share, $P_k (N_{i,t} =1\vert N_{i,t-1} =1)$. As shown in the example 
in Table III, the conditional market shares for product $B$ was always higher 
than the corresponding one for product $A$; thus, we were able to measure the 
relationship between the conditional market shares of the products by $P_k 
(N_{i,t} =1\vert N_{i,t-1} =1)$.

Fig.4 displays the relationship between the product life cycle, $T$, and the 
conditional market share, $P_k (N_{i,t} =1\vert N_{i,t-1} =1)$, in the case 
of rice balls. A linear relationship was approximated as follows:

\begin{equation}
T=20.0\cdot e^{6.04\cdot P(1\vert 1)}.
\end{equation}

\noindent
We estimated the strength of the relationship by using the cross-correlation 
function for several popular product types, in Table II. We found that the 
product life cycle is more strongly correlated with the conditional market 
share of the product, $P_k (N_{i,t} =1\vert N_{i,t-1} =1)$, than with its 
market share without any conditions, $P_k (N_{i,t} =1)$.

In Fig.5, we illustrate the probability density function of the product life 
cycle for each conditional market share, $P_k (N_{i,t} =1\vert N_{i,t-1} 
=1)$. For example, the solid line in Fig.5(a) represents the probability 
density function of rice balls when $0.07<P_k (N_{i,t} =1\vert N_{i,t-1} 
=1)\le 0.17$. We found that the function can be approximated by an 
exponential distribution, and the coefficient of exponential distribution 
depends on the conditional market share. This result suggests that we can 
model the life cycle through a Poisson process where a parameter depends on 
the conditional market share.

We discuss why the life cycle of a product depends on the conditional market 
share that expresses its repeated purchase. It is well known that the 
expenses incurred when manufacturers try to attract new customers are much 
higher than the cost of maintaining loyal customers. For example, product 
manufacturers often end up spending large amounts of money to attract new 
customers. Therefore, products that have a low probability of being 
consistently chosen by customers may quickly face termination.

\begin{figure}[htbp]
\centerline{\includegraphics[width=3.432in,height=2.464in]{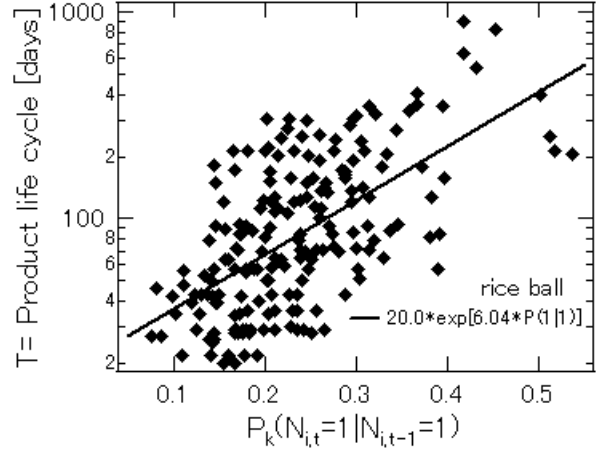}}
\label{fig4}
\caption{Relationship between product life cycle, $T$, and the 
conditional market share, $P_k (N_{i,t} =1\vert N_{i,t-1} =1)$, in the case 
of rice balls. The line represents an exponential function, $T=20.0\cdot 
e^{6.04\cdot P(1\vert 1)}$.}
\end{figure}

\begin{figure}[htbp]
\centerline{\includegraphics[width=3.447in,height=2.196in]{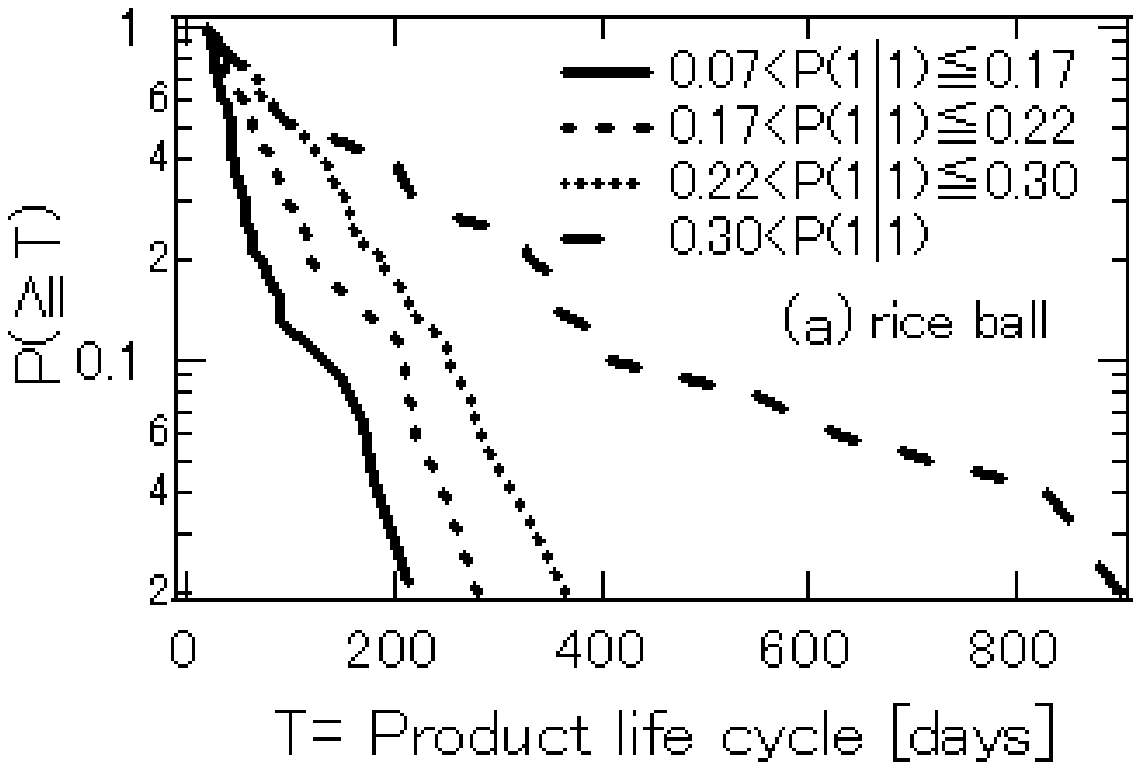}}
\label{fig5a}
\centerline{\includegraphics[width=3.44799in,height=2.09664in]{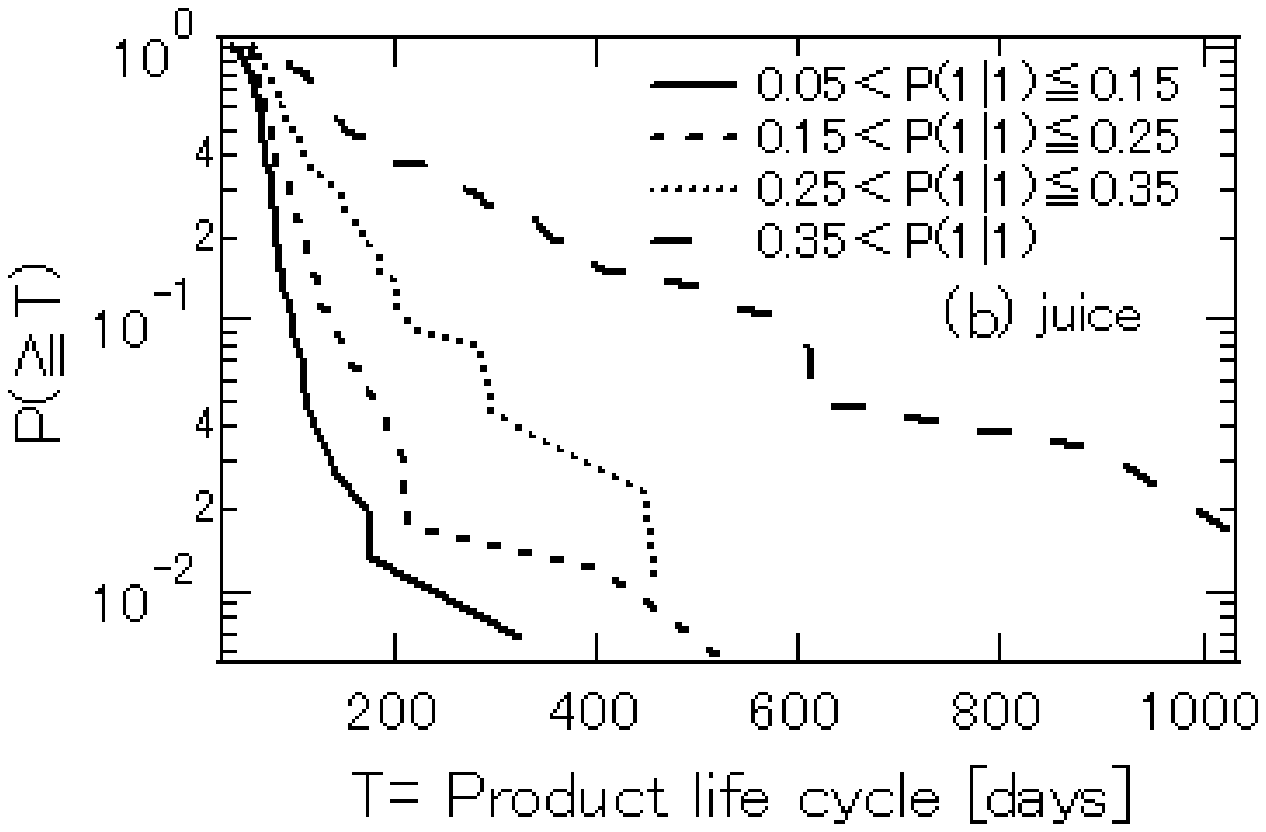}}
\label{fig5b}
\caption{Cumulative probability density functions of the product life 
cycle for each conditional market share, $P_k (N_{i,t} =1\vert N_{i,t-1} 
=1)$. (a) denotes rice balls, and (b) plastic bottles of juice. The lines 
represent the function for each value of $P_k (N_{i,t} =1\vert N_{i,t-1} 
=1)$.}
\end{figure}

\clearpage

\section{Conclusion}

In this study, we observed that the probability density function of a 
product life cycle follows the Weibull distribution or an exponential 
distribution that is a special case of the Weibull distribution. We 
introduced the conditional market share---defined as the probability that a 
product is selected by customers only if it had been purchased 
previously---and the market share without any conditions. On the basis of 
this, we found that the product life cycle is more strongly correlated with 
the conditional market share of the product than with its market share 
without conditions. Thus, we may be able to predict a product life cycle on 
the basis of its conditional market share immediately after its release 
date. We expect that these results will be applied to retail management in 
the future.

\section*{Acknowledgements}

We would like to thank Hideki Takayasu for the many discussions at various 
stages of this work. The authors also appreciate the cooperation of am/pm 
Japan Co. Ltd., in providing scanner data. Further, T. Mizuno is grateful 
for the financial support extended by ``Ken Millennium Corporation.''

%

\end{document}